\documentclass[oldversion]{aa}

\pdfoutput=1

\usepackage{txfonts} \usepackage{graphicx} \usepackage{natbib}
\bibpunct{(}{)}{;}{a}{}{,}

\newcommand{\gqcm}{\ensuremath{\mathrm{g} \, \mathrm{cm}^{-3}}}
\newcommand{\msol}{\ensuremath{M_\odot}}
\newcommand{\nhe}{\ensuremath{^{4}\mathrm{He}}}
\newcommand{\ncarb}{\ensuremath{^{12}\mathrm{C}}}
\newcommand{\nox}{\ensuremath{^{16}\mathrm{O}}}
\newcommand{\nsi}{\ensuremath{^{28}\mathrm{Si}}}
\newcommand{\nni}{\ensuremath{^{56}\mathrm{Ni}}}

\begin{document}

\title{Double-detonation supernovae of sub-Chandrasekhar mass white dwarfs}

\author{M.~Fink \and W.~Hillebrandt \and F.~K.~R\"opke}

\institute{Max-Planck-Institut f\"ur Astrophysik,
  Karl-Schwarzschild-Str.~1, D-85741 Garching, Germany}

\date{Received 7 August 2007 / Accepted 3 October 2007}

\abstract{
  Type~Ia supernovae are believed to be white dwarfs disrupted by a
  thermonuclear explosion.  Here we investigate the scenario in which
  a rather low-mass, carbon-oxygen (C~+~O) white dwarf accumulates
  helium on its surface in a sufficient amount for igniting a
  detonation in the helium shell before the Chandrasekhar mass is
  reached.  In principle, this can happen on white dwarfs accreting
  from a non-degenerate companion or by merging a C~+~O white dwarf
  with a low-mass helium one.  In this scenario, the helium detonation
  is thought to trigger a secondary detonation in the C~+~O core.  It
  is therefore called the ``double-detonation sub-Chandrasekhar''
  supernova model.

  By means of a set of numerical simulations, we investigate the
  robustness of this explosion mechanism for generic 1-\msol\ models
  and analyze its observable predictions.  Also a resolution
  dependence in numerical simulations is analyzed.

  Hydrodynamic simulations of the double-detonation sub-Chandrasekhar
  scenario are conducted in two and three spatial dimensions.  The
  propagation of thermonuclear detonation fronts, both in helium and
  in the carbon-oxygen mixture, is computed by means of both a
  level-set function and a simplified description for nuclear
  reactions.  The decision whether a secondary detonation is triggered
  in the white dwarf's core or not is made based on criteria given in
  the literature.

  In a parameter study involving different initial flame geometries
  for He-shell masses of 0.2 and 0.1~\msol\ (and thus 0.8 and
  0.9~\msol\ of C~+~O), we find that a secondary detonation ignition
  is a very robust process.  Converging shock waves originating from
  the detonation in the He shell generate the conditions for a
  detonation near the center of the white dwarf in most of the cases
  considered.  Finally, we follow the complete evolution of three
  selected models with 0.2~\msol\ of He through the C/O-detonation
  phase and obtain \nni-masses of about 0.40 to 0.45~\msol.

  Although we have not done a complete scan of the possible parameter
  space, our results show that sub-Chandrasekhar models are not good
  candidates for normal or sub-luminous type~Ia supernovae.  The
  chemical composition of the ejecta features significant amounts of
  \nni\ in the outer layers at high expansion velocities, which is
  inconsistent with near-maximum spectra.}


\maketitle

 \section{Introduction}
\label{sec:int}

One of the major uncertainties in modeling type~Ia supernovae (SN~Ia)
originates from the unknown nature of the progenitor systems because
neither observations nor theoretical models are yet conclusive
\citep{Branch1995,Ruiz-Lapuente2000,Livio2000,Nomoto2003,Han2004,Napiwotzki2005,Stritzinger2006,Parthasarathy2007}.
Over the past years, most studies of thermonuclear supernova
explosions focused on models in which a thermonuclear flame is formed
by a runaway near the center of a white dwarf (WD), composed of carbon
and oxygen, once it has reached the Chandrasekhar mass ($M_\mathrm{Ch}
\sim 1.4~\msol$) by accretion from a non-degenerate companion.
Starting out in the sub-sonic deflagration mode of flame propagation,
these models were shown to give rise to energetic explosions either in
pure turbulence-boosted deflagrations
\citep{Reinecke2002b,Gamezo2003,Roepke2005,Roepke2006a,Schmidt2006a,Schmidt2006b,Roepke2007c}
or with a delayed triggering of a supersonic detonation phase
\citep{Gamezo2004,Plewa2004,Roepke2007a}.  There seems to be
reasonable agreement of such models with the gross features of
observed SNe~Ia \citep{Roepke2007c,Mazzali2007}.  However, it remains
unclear whether the Chandrasekhar mass model can account for the full
range of SN~Ia observations.  In particular, the mechanism of the
sub-luminous events (\object{SN~1991bg} being a prototypical example)
remains a puzzle \citep{Mazzali2007}.  \citet{Stritzinger2006} even
claim typical ejecta masses below $M_\mathrm{Ch}$ for a sample of well
observed SNe~Ia, with the trend that the low-luminosity explosions
eject less mass.

Therefore, a long standing question is whether other progenitor
channels contribute to the observed SN~Ia sample, and two alternatives
have been suggested: the \emph{WD merger} or \emph{double degenerate
  scenario} (in contrast to single degenerate models, which consist of
binaries with only one WD) and the \emph{sub-Chandrasekhar model}.  In
the present work we explore the latter.

The basic idea of the sub-Chandrasekhar model is that if the accretion
rate onto a WD is lower than about $1$ - $4 \cdot 10^{-8}~\msol \,
\mathrm{yr}^{-1}$ the accreted He (or the H processed to He) is not
fused steadily into C and O.  Instead, after reaching a critical
amount of He at relatively low densities, the He shell becomes
unstable and detonates \citep[cf.][and references
therein]{Woosley1986}.  The detonation ignites most likely close to
the bottom of the He shell, produces almost pure \nni, and can occur
long before the WD reaches the Chandrasekhar limit.  A second
detonation may be triggered spontaneously when the He shell detonation
shock wave hits the C~+~O core or with some delay after the shock has
converged near the center of the WD (\emph{double detonation} models).
This way, in the sub-Chandrasekhar mass models the conditions for
thermonuclear runaway are caused by the compression in the shock wave
and not by the high degree of degeneracy as in the Chandrasekhar mass
models.

Delayed double detonations have been studied extensively before.
\citet{Woosley1994}, and \citet{Livne1990} carried out one-dimensional
(1D) simulations and \citet{Livne1990a,Livne1991} considered
two-dimensional (2D) setups.  In 1D the He detonation ignites
synchronously in a layer close to the bottom of the He shell and due
to the spherical symmetry constraint the shock converges perfectly in
the center.  However, this is not a very realistic model.  According
to \citet{Livne1990} an ignition most likely happens in a single point
leading to an off-center convergence of oblique shock waves in the
core.

Several other simulations predict successful directly ignited double
detonations in two dimensions
\citep{Livne1997,Arnett1997,Wiggins1997,Wiggins1998}.  The smoothed
particle hydrodynamics simulations by \citet{Benz1997} and
\citet{Garcia-Senz1999} were carried out in 3D\@.  \citet{Livne1990a}
first reported an increased probability of the direct core ignition if
the He detonation does not happen directly at the core--shell
interface but at a certain distance above it.  This way, the pressure
jump can grow large enough before hitting the core.

In this work, the results of a (restricted) parameter study are
presented investigating the possibility of triggering the second
detonation in two and three dimensions.  In all our models the WD has
a total mass of 1~\msol, but the (C~+~O)-core and the He-shell masses
differ.  We compute sequences of models with very different initial
flame geometries in order to test the robustness of this explosion
mechanism.  Since this latter question is the main focus of our paper,
most of the simulations are stopped once the conditions for a
detonation are matched.  However, for a few successful cases the
energetics and nucleosynthesis of complete double detonations are
computed.  In the following chapter the model details will be
described.  Section~\ref{sec:sim} shows the simulation results, which
are summarized and discussed in the last part of the paper.

\section{Explosion model}
\label{sec:mod}

\subsection{General setup}
\label{sec:setup}

The numerical scheme used in our simulations is a modified version of
the code described by \citet{Reinecke1999a,Reinecke2002a},
\citet{Roepke2005}, \citet{Golombek2005}, and \citet{Roepke2007a}.
The main difference is the burning physics which accounts for He
detonations and the propagation of two detonation waves.

Our hydrodynamic simulations were carried out on a uniform Eulerian
grid with a finite volume scheme based on the PROMETHEUS
implementation \citep{Fryxell1989} of the ``piecewise parabolic
method'' (PPM) \citep{Colella1984}.  Thus, discontinuities like shocks
were automatically and accurately captured but, nonetheless, smeared
out over several grid cells.  A potential concern is the overheating
effect observed in Godunov-type schemes when shocks hit walls or
collide.  Following the suggestion of \citet{Donat1996}, an
approximate Riemann solver (the so-called Marquina solver) was
implemented but this did not lead to noticeable differences.  Thus,
all simulations presented in the following are based on the exact
Riemann solver of \citet{Colella1985}.

While the parameter study testing the possibility of igniting a
detonation in the C~+~O core was carried out on a static grid, the
complete double detonation models presented at the end of
Sect.~\ref{sec:sim} employed a co-expanding grid as in
\citet{Roepke2005} in order to account for the expansion of the WD\@.
The coordinates were $(r, z)$-cylindrical in 2D and cartesian in 3D\@.

The equation of state we used includes contributions from an
arbitrarily degenerate and arbitrarily relativistic electron gas, a
photon gas, electron--positron pairs and nuclei with a
Maxwell-Boltzmann distribution.  The initial WD was assumed to be in
hydrostatic equilibrium.  Therefore its structure is uniquely
determined once the density at the center and at the core--shell
interface and the temperature profile are fixed.  For simplicity a
constant temperature of $5 \cdot 10^5~\mathrm{K}$ was assumed
throughout the star.  Of course, this is an unrealistic choice, but
because of the high degree of degeneracy the exact temperature has
almost no effect on the state of the matter.  Another simplifying
assumption is the approximation of the gravitational potential in
spherical symmetry.  Only the monopole moment of the density
distribution averaged over all angles was calculated at every time
step.

\subsection{Nuclear reactions}
\label{sec:nuclreact}

As we are mostly interested in the hydrodynamic evolution of the
explosion, it is appropriate to use a strongly simplified scheme for
nuclear reactions as in \citet{Reinecke2002a}.  On our numerical grid
the structure of the flame cannot be resolved.  Therefore in our
simulations the detonation flame is approximated as a discontinuity
propagating at an appropriate speed (see Sect.~\ref{sec:det}) with the
fast reactions happening spontaneously at its passage and only the
density dependent products, but no reaction rates being calculated.
In order to approximate the energy release in the fast reactions and
the correct molecular weight of fuel and ashes, it is sufficient to
consider only one representative species for each group of elements
involved which has a binding energy and mass typical for the group.
Our scheme includes four species, namely \nhe, a mixture of \ncarb\ 
and \nox\ with equal mass fractions, \nsi\ as a representative for the
intermediate mass elements, and \nni\ as an iron group element.  The
following reactions are included in the model, depending on the
density at the front:
\begin{itemize}
\item The \nhe\ detonation in the shell produces pure \nni.  Flame
  extinction is assumed to occur at $\rho = 10^5~\gqcm$.
\item A detonation in the \ncarb/\nox\ core produces \nni\ in nuclear
  statistic equilibrium (NSE) (represented by a temperature and
  density dependent mixture of \nni\ and \nhe) above $\rho = 3 \cdot
  10^7~\gqcm$ and \nsi\ below.\footnote{According to
    \citet{Imshennik1984}, at densities lower than about $10^7~\gqcm$
    the timescale for NSE formation becomes so large that the flame
    width would be bigger than the extent of the whole core.  Since it
    can be assumed that NSE is no longer reached even before the flame
    width gets close to the extension of the core, $3 \cdot
    10^7~\gqcm$ instead of $10^7~\gqcm$ is used as the transition
    density.}  At densities smaller than $10^6~\gqcm$ burning is
  assumed to cease.
\end{itemize}
Since the transition density to NSE affects the \nni-mass and thus the
total explosion energy, its exact value should be investigated further
in the future.

In order to distinguish the \nhe\ in the shell from that produced in
NSE with \nni\ both are treated as different species in the code.
With decreasing densities the \nhe\ in NSE finally recombines to \nni.
This is the only slow reaction included in our simulations.

The propagation of the detonation flame is modeled with the
\emph{``level set technique''} \citep{Osher1988} in the implementation
of \citet{Golombek2005}, which in turn is based on the deflagration
flame model of \citet{Reinecke1999b}.  The zero level set of a signed
distance function $G$ represents the flame front.  Information on
whether matter has already been burnt is encoded in its sign.  It is
negative in the fuel and positive in the ashes.  Most important, the
location of the flame can be calculated at every time step simply by
finding the roots of $G$.  The propagation of the zero level set of
$G$ consists of two parts: the advection with the flow and
self-propagation by burning, respectively.  By means of an operator
splitting, the advection of $G$ can be treated in the same way as that
of a passive scalar in the hydrodynamic scheme
\citep{Mulder1992,Reinecke1999a,Reinecke1999b}.  Burning is accounted
for by \citep{Reinecke2001}
\begin{displaymath}
  \left( \frac{\partial G}{\partial t} \right)_\mathrm{burn} = 
  s \left| \vec{\nabla} G \right|.
\end{displaymath}
Thus the velocity $s$ of the flame relative to the flux has to be
determined at every time step from the quantities on the grid.  In the
state-of-the-art deflagration codes this is done by an appropriate
sub-grid turbulence model.  For detonations, as occurring in our
models, turbulence effects can be neglected and the propagation speed
$s$ can be determined easily as a function of the density $\rho$ only,
as will be described in the next section.

The changes in the mass fractions are calculated depending on $\rho$,
when the flame front passes a cell.  Here and also in the case of the
velocity $s$ we have the problem that in general the state in such a
cell is a mixture of a burnt and an unburnt matter.  So we have
e.g.\footnote{Here and in the following indices 1 and 2 refer to
  quantities on the unburnt and the burnt side of the front,
  respectively.}:
\begin{displaymath}
  \rho = \alpha \rho_1 + (1 - \alpha) \rho_2,
\end{displaymath}
where $\alpha$ is the volume fraction of the cell occupied by the
unburnt state.  With the so called \emph{``complete coupling''} scheme
the two partial states could be reconstructed from the mixed state
\citep{Smiljanovski1997}.  This enables flux splitting that avoids
numerical mixing of fuel and ashes.  While such a scheme has been
implemented for 2D small-scale deflagration flame simulations in WD
matter \citep{Roepke2003,Roepke2004a,Roepke2004b}, numerical obstacles
prevent its application in full star simulations up to now.  Here, we
apply the ``passive implementation'' of \citet{Reinecke1999b} and use
the mixed state to advect $G$ and also to calculate the burning
velocity $s$ and to determine the reaction
products.\footnote{Correctly, one would have to use burning and flow
  velocities on either the burnt or the unburnt side of the flame to
  update $G$ due to burning and advection.}

The scheme described here is not in general appropriate to simulate
detonations as, e.g., the burning speed $s$ depends on the detailed
geometry of the detonation front.  The burning speed $s$ is in general
a function of $\frac{\delta}{R}$, where $\delta$ is the width of the
detonation front and $R$ is its curvature radius.  In our case,
however, the front can be approximated as thin and planar in the
density range that is relevant for the explosion dynamics.

\subsection{Detonation physics}
\label{sec:det}

A detonation wave consists of a shock wave that compresses and heats
the matter behind it so strongly that it is ignited.  The shock wave,
in turn, is maintained by the energy released due to burning.
Depending on the properties of the reactions taking place, there exist
different types of detonations.  A very important case is the
\emph{Chapman-Jouguet detonation}, which propagates into the fuel at
the lowest possible speed.  According to \citet{Landau1991} a
detonation that is created spontaneously by the combustion itself is a
Chapman-Jouguet detonation in many cases.  Therefore, in our
simulations we assumed this type of detonation for the He burning.  In
the Chapman-Jouguet case there exists a simple law for the velocity of
the burning front relative to the ashes, which is just equal to the
sound speed there:
\begin{displaymath}
  s_\mathrm{2,He} = \sqrt{\gamma_2 
    \left( \frac{\partial p_2}{\partial \rho_2} \right)_T}.
\end{displaymath}
Here $\gamma = \frac{c_p}{c_V}$ is the ratio of the specific heat at
constant pressure and the specific heat at constant volume.

According to \citet{Sharpe1999}, self supporting detonations in
degenerate carbon/oxygen mixtures at densities between $2 \cdot 10^7$
and $1 \cdot 10^9~\gqcm$ are of \emph{pathological} type.  These arise
in general from a non-monotonic heat release in the reactions.  For
them the front velocity cannot be determined independent of the
reaction rates.  Therefore, for the C/O detonations in this work, we
used the values calculated by \citet{Sharpe1999} in numerical
simulations as a function of density:
\begin{displaymath}
  s_\mathrm{1,C/O} = f_\mathrm{Sharpe}(\rho).\footnotemark
\end{displaymath}%
\footnotetext{We neglect here that over-driven detonations
  powered by shock waves entering the core due to the He shell
  detonation could also play a role.}

\subsection{Ignition conditions in the C~+~O core}
\label{sec:ign_cond}

As our code lacks a detailed nuclear network and reaction rates are
not calculated, the onset of a C/O detonation can not be followed.
Apart from that, in our full star simulations it would not be feasible
to resolve the scales of the detonator (where the shock that supports
the detonation is built up).  Therefore, the decision on whether a
detonation is triggered is made upon critical conditions given by
\citet{Niemeyer1997} and \citet{Roepke2007b} (see
Tables~\ref{tab:ignhigh} and \ref{tab:ignlow}).
\begin{table}
  \caption{Critical masses for detonations at given density and
    temperature according to \citet{Niemeyer1997}.}
  \label{tab:ignhigh}
  \centering
  \begin{tabular}{rccr}
    \hline
    \hline
    $\rho~[\gqcm]$ & $T_\mathrm{c}~[10^9~\mathrm{K}]$ &
    $M~[\mathrm{g}]$ & $R~[\mathrm{m}]$ \\
    \hline
    $2 \cdot 10^9$ & 2.8 & $2.0 \cdot 10^{15}$ & 0.7 \\
    $10^8$         & 3.2 & $2.0 \cdot 10^{15}$ & 2   \\
    $3 \cdot 10^7$ & 3.2 & $2.0 \cdot 10^{19}$ & 50  \\
    \hline
  \end{tabular}
\end{table}
\begin{table}
  \caption{Critical temperatures for detonations at given density and
    mass according to \citet{Roepke2007b}.  Detonation is impossible
    for  $\rho \le 1 \cdot 10^6~\gqcm$.}
  \label{tab:ignlow}
  \centering
  \begin{tabular}{rccr}
    \hline
    \hline
    $\rho~[\gqcm]$ & $T_\mathrm{c}~[10^9~\mathrm{K}]$ &
    $M~[\mathrm{g}]$ & $R~[\mathrm{km}]$ \\
    \hline
    $10^7$       & 2.8 & $2.5 \cdot 10^{23}$ & 2   \\
    $10^7$       & 2.0 & $2.0 \cdot 10^{25}$ & 8   \\
    $10^7$       & 1.9 & $1.5 \cdot 10^{27}$ & 30  \\
    $3\cdot10^6$ & 2.3 & $2.0 \cdot 10^{28}$ & 120 \\
    \hline
  \end{tabular}
\end{table}
They are based on the following arguments.  If there is a sufficiently
large region of supersonic spontaneous burning, a strong pressure wave
must form and will synchronize with the burning wave if the
compression and heating is strong enough to burn the material in the
sound crossing time of the region under consideration
\citep{Blinnikov1986,Woosley1990,Khokhlov1991b}.  Therefore a
detonation wave is born if that region is big enough or, equivalently,
provided that it consists of enough mass at a given density.
\citet{Niemeyer1997} empirically determined those \emph{critical
  masses} from hydrodynamic simulations for different densities and
for a given linear\footnote{with respect to inner mass} radial
temperature gradient across that mass.\footnote{In
  Table~\ref{tab:ignhigh} only the results for a mixture of equal mass
  fractions of C and O are shown.  In their work
  \protect{\citet{Niemeyer1997}} additionally investigated the
  dependence on the mass fraction.}  \citet{Roepke2007b} used the same
code and setup, but determined the critical temperature at a given
mass for different densities.  They focussed on the low end of
possible densities.

\section{Simulations}
\label{sec:sim}

With the numerical scheme described above, several simulations were
performed.  In Sect.~\ref{sec:fla}, the results of a parameter study
on the influence of the initial flame geometry for two different core
and He-shell masses are presented, followed by a resolution study in
Sect.~\ref{sec:res} and, finally, the determination of explosion
energies and nucleosynthesis yields in three successful double
detonation simulations.  We stress again that the main intension of
the parameter study is not to scan the entire parameter space of
possible WD properties but rather to investigate the conditions which
may lead to a double detonation.

\subsection{Initial flame geometry}
\label{sec:fla}

The accretion of matter from the companion star onto the WD is a very
complex process, and the formation and ignition of a He shell is still
afflicted with large uncertainties.  As many different initial flame
geometries in the shell seem possible, their influence on the
possibility and the properties of a double detonation is explored
first.

The different initial flame models, which were treated in this study,
are shown in Fig.~\ref{fig:inimod} and the nomenclature of the
simulations is as follows:
\begin{figure*}
  \includegraphics[width=12cm]{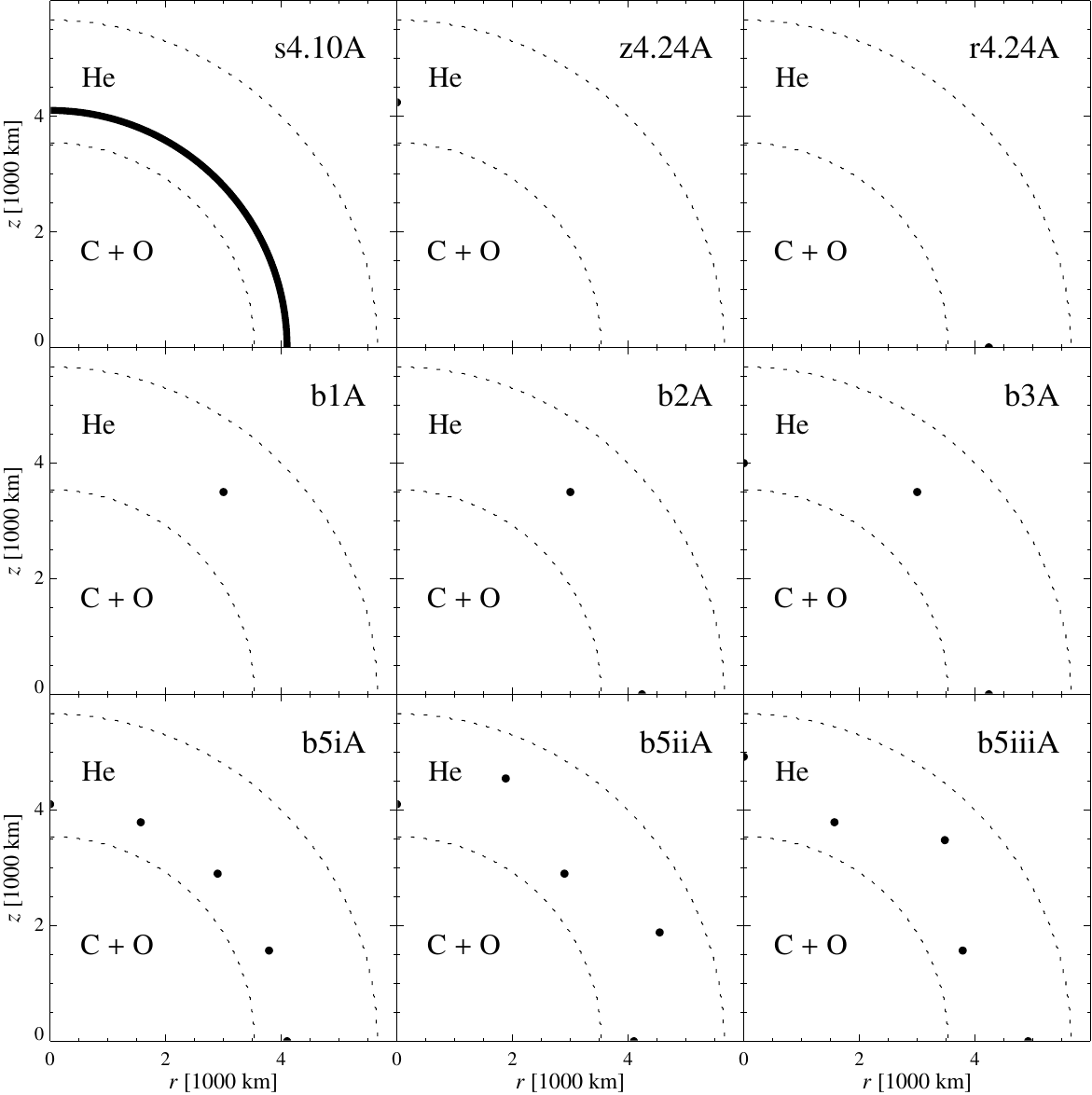}
  \caption{Initial flame models.  The He detonation flame fronts start
    at the borders of the black regions that already contain hot ashes
    at the beginning.}
  \label{fig:inimod}
\end{figure*}
The first letter in each model's name denotes the geometry, and the
following number specifies it further.  ``s4.10'' means an ignition in
a whole layer of the He shell at a radius of 4100~km.  ``b$n$'' stands
for an ignition in $n$ separate ``bubbles'' (which are, of course,
tori in 2D if located off the z-axis).  Two special cases of the
latter type, for $n = 1$, are ``z5.00'' and ``r5.00'', where the
initial ``bubble'' is located at the $z$- or the $r$-axis,
respectively, at a distance of 5000~km from the origin.  Finally the
capital letters ``A'' or ``B'', which follow the above names, indicate
the initial masses which were used:
\begin{itemize}
\item Model A: $\quad M_\mathrm{C/O} = 0.8~\msol, \quad M_\mathrm{He}
  = 0.2~\msol$,
\item Model B: $\quad M_\mathrm{C/O} = 0.9~\msol, \quad M_\mathrm{He}
  = 0.1~\msol$.
\end{itemize}

The results of the study are given in Table~\ref{tab:flmod}.
\begin{table}
  \caption{Results of the parameter study about the initial flame
    geometry.  The last column states if the ignition conditions of a
    core detonation have been reached.}
  \label{tab:flmod}
  \centering
  \begin{tabular}{llllc}
    \hline
    \hline
    Model  &  $t/{\mathrm s}$  &  $T_{\mathrm 9,\,max}$  &
    $\rho_{\mathrm 8,\,max}$  &  Core ignition \\
    \hline
    s3.60A\_2dq\_256 & 0.627 & 9.09 & 6.14  & yes \\
    s4.10A\_2dq\_256 & 0.575 & 10.2 & 7.37  & yes \\
    s5.00A\_2dq\_256 & 0.609 & 12.1 & 9.05  & yes \\
    r4.24A\_2dq\_256 & 0.699 & 5.62 & 1.54  & yes \\
    r5.00A\_2dq\_256 & 0.718 & 6.00 & 1.69  & yes \\
    z4.24A\_2dq\_256 & 0.989 & 3.27 & 1.03  & no  \\
    z5.00A\_2dq\_256 & 1.03  & 3.20 & 0.981 & no  \\
    b1A\_2dq\_256    & 0.767 & 4.43 & 1.32  & yes \\
    b2A\_2dq\_256    & 0.678 & 8.61 & 4.85  & yes \\
    b3A\_2dq\_256    & 0.674 & 7.94 & 4.57  & yes \\
    b5iA\_2dq\_256   & 0.629 & 9.53 & 6.68  & yes \\
    b5iiA\_2dq\_256  & 0.648 & 9.00 & 6.13  & yes \\
    b5iiiA\_2dq\_256 & 0.648 & 9.03 & 6.37  & yes \\
    \hline
    z4.24A\_2d\_256 & 1.08 & 8.53 & 4.52 & yes \\
    \hline
    s4.30B\_2dq\_256 & 0.791 & 6.82 & 4.69  & yes \\
    s4.80B\_2dq\_256 & 0.745 & 8.17 & 5.99  & yes \\
    r4.50B\_2dq\_256 & 0.892 & 3.96 & 1.15  & no  \\
    r5.00B\_2dq\_256 & 0.885 & 4.49 & 1.23  & yes \\
    z4.50B\_2dq\_256 & 1.21  & 2.42 & 0.808 & no  \\
    z5.00B\_2dq\_256 & 1.24  & 2.39 & 0.768 & no  \\
    b2B\_2dq\_256    & 0.855 & 6.72 & 4.34  & yes \\
    b5iB\_2dq\_256   & 0.806 & 7.63 & 5.47  & yes \\
    \hline
    z4.50B\_2d\_256 & 1.31 & 6.50 & 3.38 & yes \\
    \hline
  \end{tabular}
\end{table}
There the dimensionality and the number of grid cells along the side
of one quadrant are added to the model names.  Throughout the study a
resolution of 256 cells was used.  ``2dq'' means that only the first
quadrant has been simulated assuming equatorial symmetry, whereas
``2d'' denotes full two dimensional simulations of the whole
$z$-range, capturing the entire WD star.

The most symmetric case with the simplest hydrodynamic flow pattern is
the ignition in a whole layer of the He shell.  In model
s4.10A\_2dq\_256 starting with the initial flame geometry shown at the
top left of Fig.~\ref{fig:inimod}, the inner and the outer spherical
detonation front start to propagate inwards and outwards,
respectively.  While the combustion stops close to the border of the
WD due to the low densities, it ceases at the core--shell interface
because of the depletion of the He fuel.  To investigate the chances
of a core ignition, only the hydrodynamic evolution was considered
afterwards and all the reactions taking place were stopped.  The shock
wave that was formerly driving the inner detonation front continues to
propagate inwards until it converges at the center.  There (at $t =
0.575~\mathrm{s}$) a strong shock collision takes place that causes a
maximum temperature of $10.2 \cdot 10^9$~K and densities up to $7.37
\cdot 10^8~\gqcm$.

According to Table~\ref{tab:ignhigh} these maximum values suffice for
a C/O ignition, if there is at least a mass of about $2.0 \cdot
10^{15}~\mathrm{g}$ with the same temperature and density or, more
accurately, with a sufficiently small temperature decrease from the
above mentioned value.  A mass of $2.0 \cdot 10^{15}~\mathrm{g}$ at
this density corresponds to a spherical volume with a radius of the
order of 1~km.  This length was not resolved in this simulation, where
one cell has a diameter of $23.2~\mathrm{km}$.  But as the high
temperatures and densities in our simulations are not confined to one
single cell, but distributed over several (which means that the region
of interest is resolved), the ignition conditions seem to be clearly
fulfilled.  Consequently, a C/O detonation will occur at the center of
the core.  There is, however, the possibility that a detonation of the
C/O material is triggered already when the shock wave enters at the
edge of the core.  About 0.1~s after the start of the simulation,
temperatures of about 1 to $2 \cdot 10^9~\mathrm{K}$ and densities of
about $10^7~\gqcm$ are reached.  According to Table~\ref{tab:ignlow},
this is within the range of conditions where a detonation might occur.
But as the temperature gradient might possibly be too large across a
radius of about 30~km, this question remains open here.  As was
already mentioned, other authors
\citep{Livne1997,Arnett1997,Benz1997,Wiggins1997,Wiggins1998,Garcia-Senz1999}
found successful direct ignitions and an increasing probability for
the latter with increasing distance of the He ignition point from the
C/O core \citep[cf.][]{Livne1990a}.

Our simulations confirm the earlier results.  Helium shell detonations
were ignited at smaller (s3.60A\_2dq\_256) and larger
(s5.00A\_2dq\_256) radii than in the case described above.  As was
expected, the shock wave propagating through the core is weaker in the
case of smaller ignition radii and stronger in the case of shocks
starting further out.  Accordingly the maximum temperatures and
densities that are reached in the center are lower in the former and
higher in the latter case (see Table~\ref{tab:flmod}).

Due to the high degree of symmetry, the models with shell ignition are
the ones that can trigger a second detonation most easily.  To test
multi-dimensional configurations, the complete ignition layer was
replaced by single spots (which effectively are tori in 2D)
distributed over the He shell.  Although this is probably not a
realistic scenario, it is considered here as a numerical experiment.
As far as their maximum densities and temperatures are concerned, the
models with five initially ignited bubbles in the first quadrant
differ only slightly from the complete shell ignition model (see
Table~\ref{tab:flmod}).  The values of $T_\mathrm{max}$ and
$\rho_\mathrm{max}$ are only slightly lower than for the ignition of a
shell, starting at the same radius, and they still match the
conditions for a detonation.  If the number of initially burning
bubbles is reduced to three, peak temperature and density decrease
further.

\begin{figure*}
  \includegraphics[width=12cm]{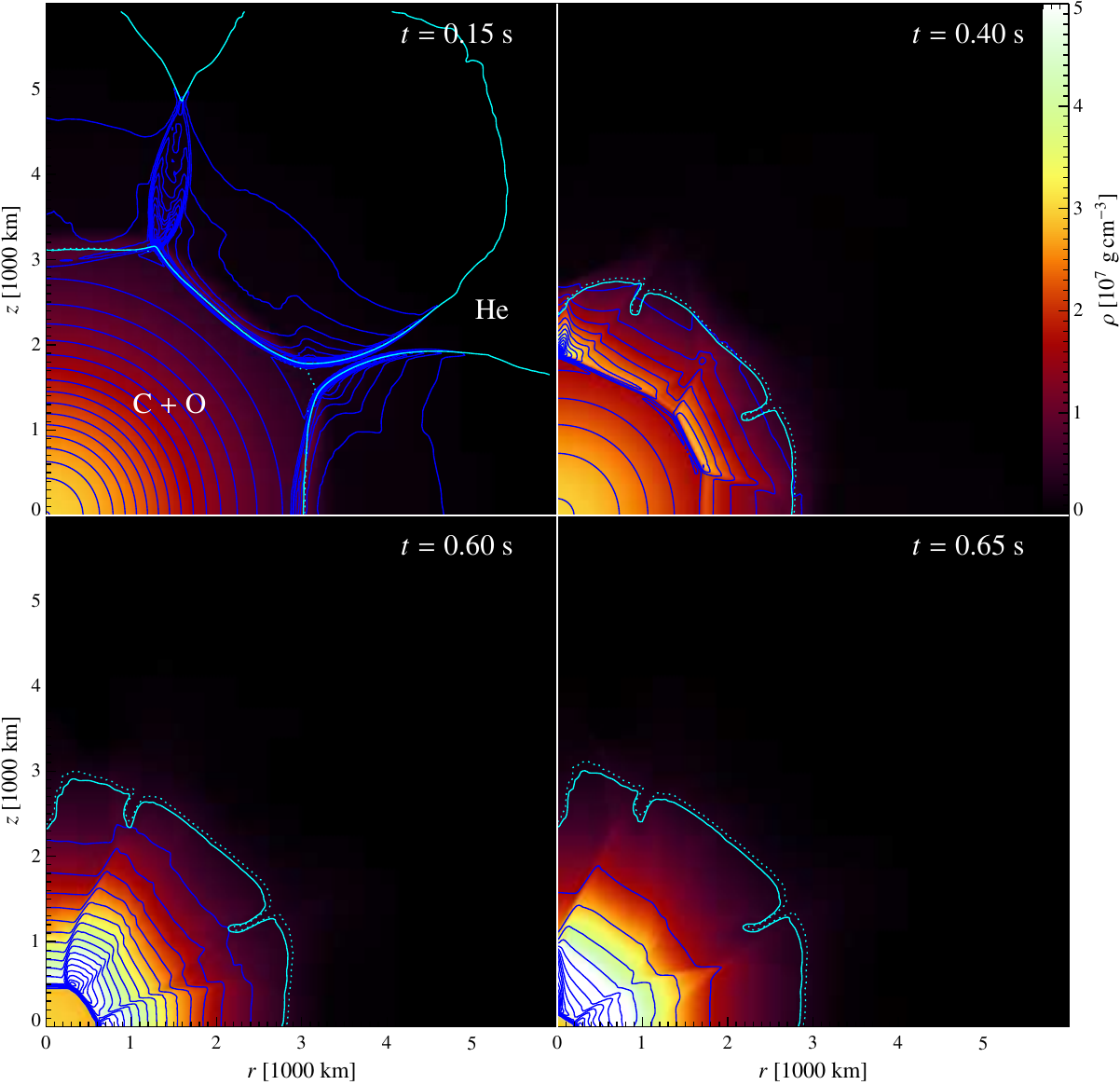}
  \caption{Time evolution of the model b3A\_2dq\_256.  Here and in the
    following figures the density is color coded and lines of constant
    pressure are drawn in blue.  In cyan the solid line is the
    location of the He detonation flame (the zero level of $G$) and
    the dashed lines are at the border of the He shell.}
  \label{fig:b3A}
\end{figure*}

The evolution of simulation b3A\_2dq\_256 is shown in
Fig.~\ref{fig:b3A}.
Due to the coalescence of the parts of the detonation fronts moving
inwards, an almost spherically symmetric shock front is entering the
core.  This resemblance of the dominant hydrodynamic processes is the
reason for the similarity of the results given in
Table~\ref{tab:flmod}.

In between the bubbles interactions of the detonation shocks take
place.  The collision process of two radial shocks can be described as
a reflection at the plane of symmetry between the waves \citep[see
e.g.][]{Courant1948}.  Generally, in this frame the incidence angle
may differ from the angle of reflection.  This can be seen best at $t
= 0.15~\mathrm{s}$ in Fig.~\ref{fig:b3A}.  The reflected waves are
again reflected at the waves coming from the opposite side of the
initial bubble.  The repetition of this process should also have some
influence on the geometry of the innermost shock front and thus on the
conditions given in Table~\ref{tab:flmod}.

This can be seen in model b2A\_2dq\_256, which is similar to model
b3A\_2dq\_256, but with the bubble on the $z$-axis removed.  It starts
with only two initially ignited bubbles but reaches greater maximum
densities and temperatures in the core.  The reason may be a focusing
effect on the $z$-axis in 2D cylindrical symmetry and the bigger
density and pressure jump that can build up until the burning reaches
the axis.

An analogous focusing effect in the plane of equatorial symmetry (the
$xy$-plane) is not observed.  On the contrary, removing the initial
burning bubble from the $r$-axis (i.e.\ going from b2A\_2dq\_256 to
b1A\_2dq\_256) leads to a strong decrease of the maximum values of
density and temperature.

\begin{figure}
  \resizebox{\hsize}{!}{\includegraphics{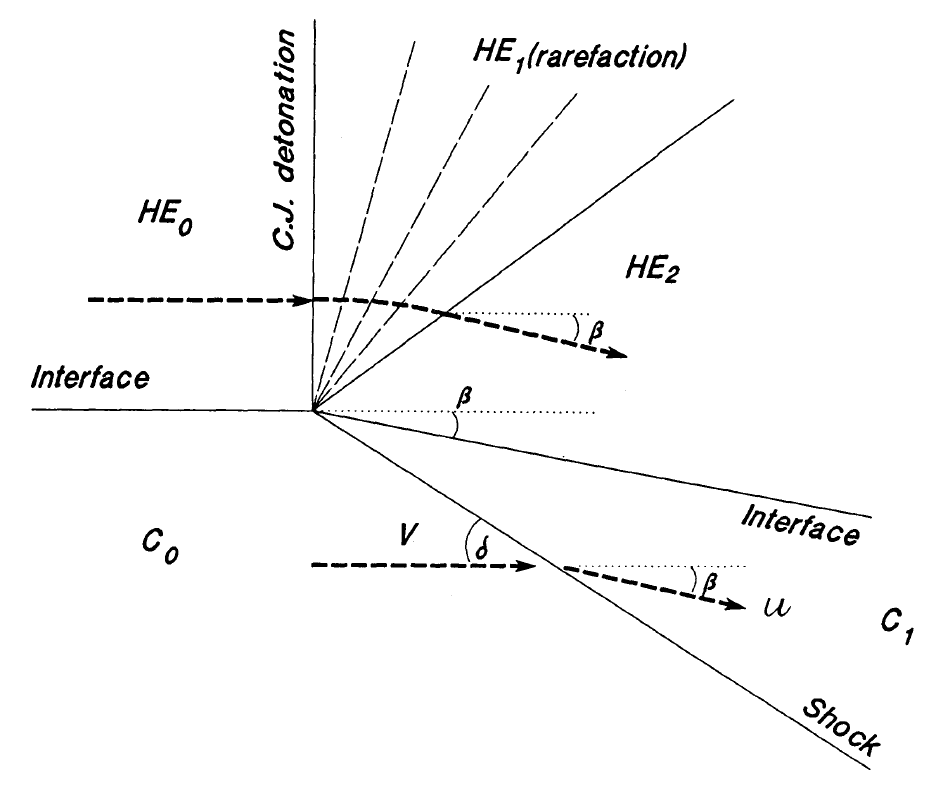}}
  \caption{Structure of a detonation propagating at the core--shell
    interface in the rest frame of the detonation front from
    \citet{Livne1990a}.  (Here the core consists of carbon only, not
    of C/O like in our case.)}
  \label{fig:oblshock}
\end{figure}

Considering models with only a single initially burning bubble, the
exact geometry of the appearing shock waves is of great importance.
After the He detonation has reached the border of the core, it spreads
around it in a wave propagating almost perpendicular to the
core--shell interface.  The detailed hydrodynamic structure has been
analyzed in \citet{Livne1990a}, from which Fig.~\ref{fig:oblshock} is
taken.
An oblique shock wave that propagates into the core (tilted by an
angle $\delta$ with respect to the interface) is associated with the
detonation wave moving along the surface of the core.  After the shock
passage the core--shell interface is also tilted inwards by an angle
$\beta$.  Thus in-flowing matter follows the oblique shock.  The
detonation wave is followed by a rarefaction wave.  The angles $\beta$
and $\delta$ can be determined by applying the Rankine-Hugoniot
conditions and depend on the strength of the detonation shock.

\begin{figure*}
  \includegraphics[width=12cm]{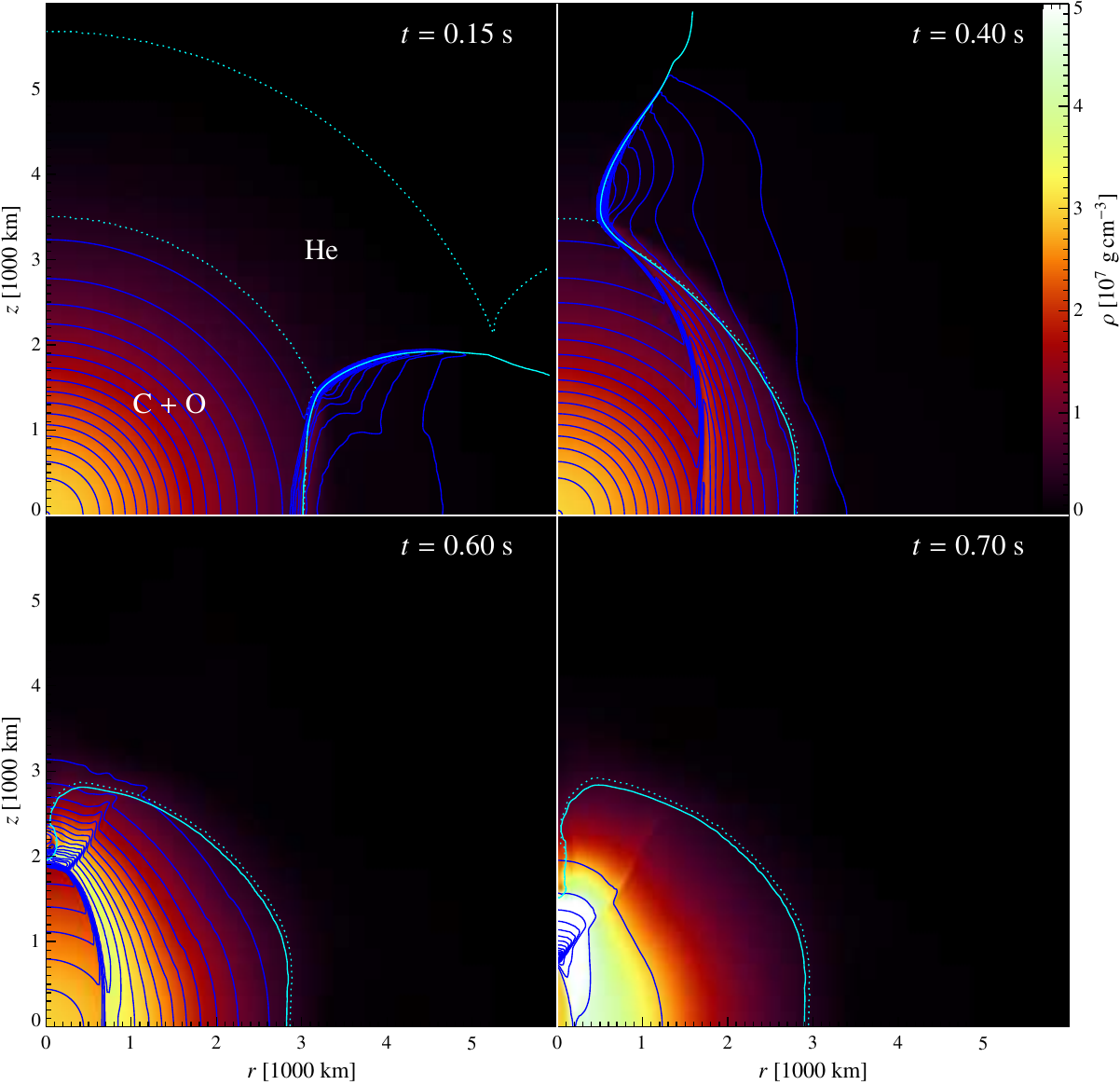}
  \caption{Time evolution of the model r4.24A\_2dq\_256.}
  \label{fig:r4.24A}
\end{figure*}

This structure can clearly be seen at $t = 0.40~\mathrm{s}$ in model
r4.24A\_2dq\_256, which is the detonation of a torus in the $z = 0$
plane (see Fig.~\ref{fig:r4.24A}).
Starting from the ``equatorial'' plane the detonation wave propagates
around the core until it reaches the ``polar'' $z$-axis.  There a
collision of the waves coming from all azimuthal directions takes
place.  Equivalently, each wave is reflected at the wave coming from
the opposite side of the star.  This causes the shock wave visible in
the last two snapshots of Fig.~\ref{fig:r4.24A} to propagate to the
lower right.  In the following the oval inner shock front moves
towards the center.  At $t = 0.70~\mathrm{s}$, even before the center
is reached, the maximum density and temperature values are attained on
the $z$-axis at about 850~km away from the center (fourth plot in
Fig.~\ref{fig:r4.24A}).  The density and temperature values of $1.54
\cdot 10^8~\gqcm$ and $5.62 \cdot 10^9~\mathrm{K}$ are significantly
lower than those in the spherically symmetric models, but according to
Table~\ref{tab:ignhigh} they still suffice for the ignition of an
off-center core detonation.

\begin{figure}
  \resizebox{\hsize}{!}{\includegraphics{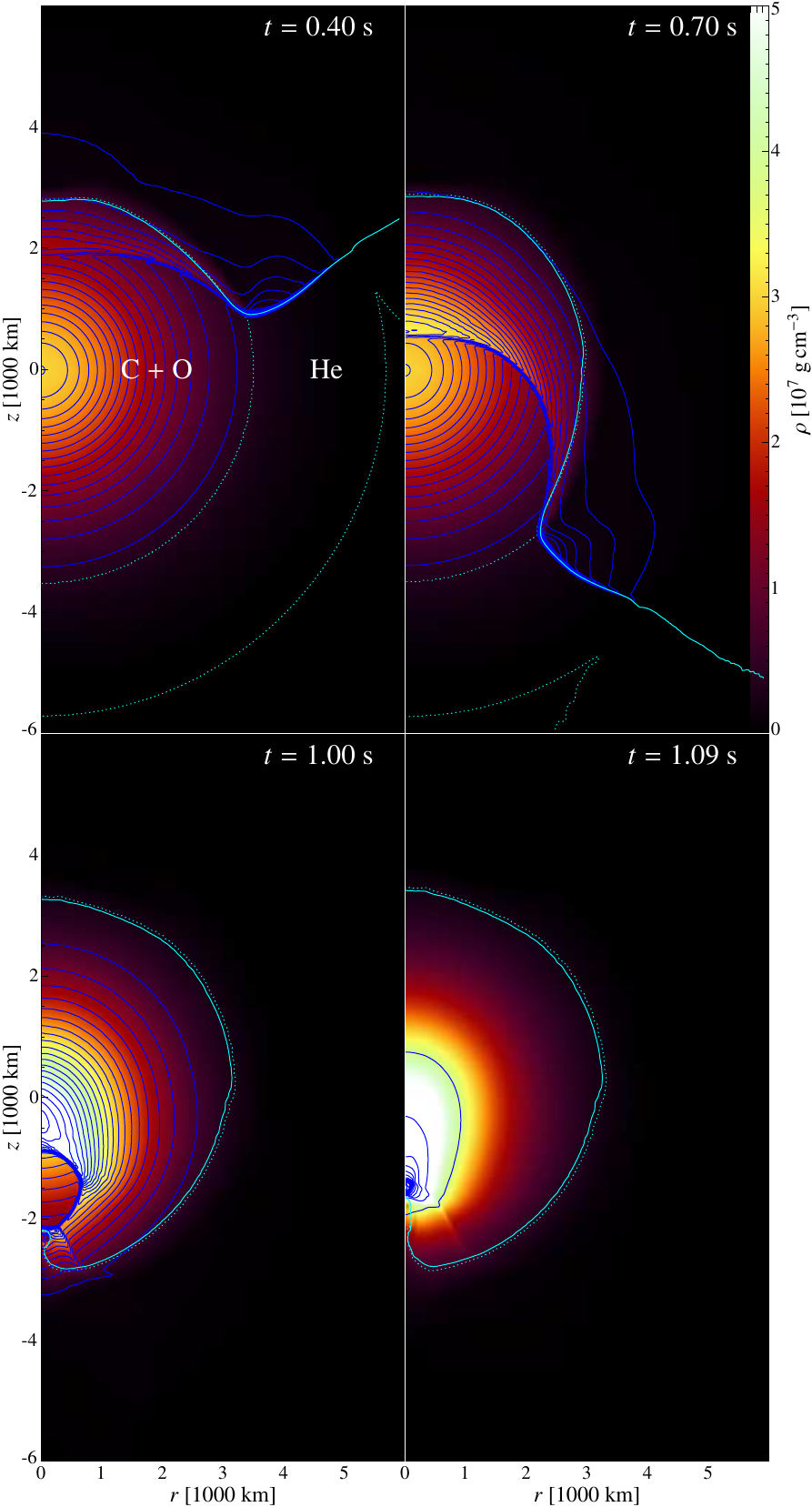}}
  \caption{Time evolution of the model z4.24A\_2d\_256.}
  \label{fig:z4.24A}
\end{figure}

In the following simulations, the symmetry of the first detonation was
further reduced by departing from equatorial symmetry and simulating a
whole rotationally symmetric star, i.e.\ now the negative $z$-range
was considered too.  Fig.~\ref{fig:z4.24A} shows the example
simulation z4.24A\_2d\_256.  Here the He detonation starts in a point
on the $z$-axis and then spreads around the whole WD until it reaches
the negative $z$-axis (see Fig.~\ref{fig:z4.24A}).  In this sense, the
setup is maximally asymmetric.
The processes now taking place are analogous to the simulation
r4.24A\_2dq\_256 described above.  But here the maximum density and
temperature are reached when the oval inner shock front converges
almost perfectly in a ``point'' on the negative $z$-axis at $z \approx
-1500~\mathrm{km}$.  Thus at a time of $1.08~\mathrm{s}$,
$\rho_\mathrm{max} = 4.52 \cdot 10^8~\gqcm$ and $T_\mathrm{max} = 8.53
\cdot 10^9~\mathrm{K}$ are reached.  These maximum values are
comparable with the spherically symmetric case discussed at the
beginning of this section.  The reason for this is the high degree of
symmetry of the colliding shock fronts.  Astoundingly those maximum
values are reached comparatively far away from the center in a region
of relatively low initial density.

The initial condition considered in the last paragraph seems more
realistic than the more symmetric cases.  Thus, this model was studied
also in simulations with higher numerical resolution.  Here, a
resolution dependence of the peak temperatures and densities became
apparent -- an effect that is discussed in Sect.~\ref{sec:res} below.

Finally, a series of models with a different initial mass distribution
was investigated.  Sequence ``B'' corresponds to models with
$0.9~\msol$ of C/O and $0.1~\msol$ of He and the parameters given in
Table~\ref{tab:flmod}.  They support the generality of the results
discussed above.  The dependence on the initial flame geometry is
analogous to case ``A'', but maximum densities and temperatures
achieved are generally lower.  This is consistent with the fact that
the shock waves originating from the He shell have to cross a larger
pressure gradient before they reach the center in model
``B''.\footnote{Geometrical amplification effects do not play a role
  at such big radii.}  Still, most of the models reach the conditions
for core ignition described in Sect.~\ref{sec:ign_cond} (the ``z$r$''
models did not ignite with mass distribution ``A'' either).

\subsection{A resolution study}
\label{sec:res}

\begin{figure}
  \resizebox{\hsize}{!}{\includegraphics{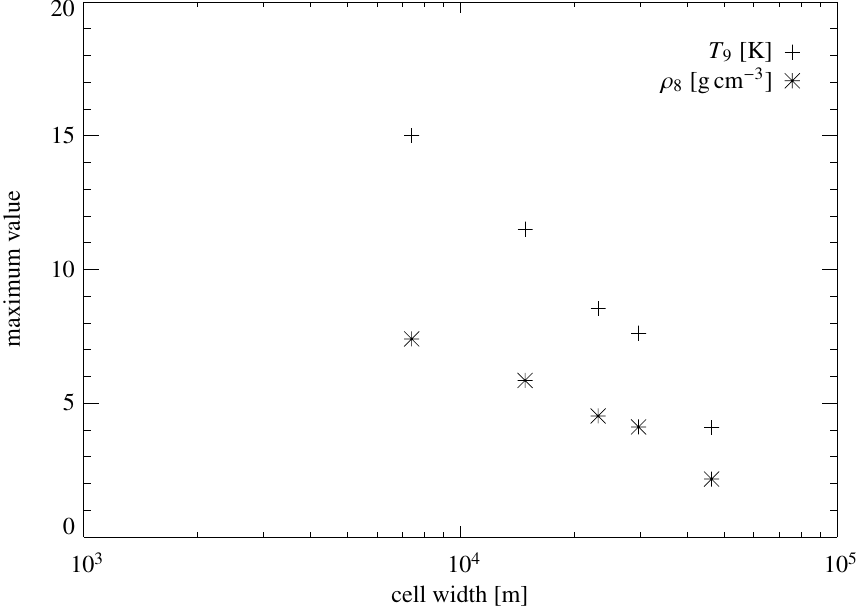}}
  \caption{Resolution study of the maxima in the central shock
    collision of model z4.24A\_2d.}
  \label{fig:resstudy}
\end{figure}

\begin{figure}
  \resizebox{\hsize}{!}{\includegraphics{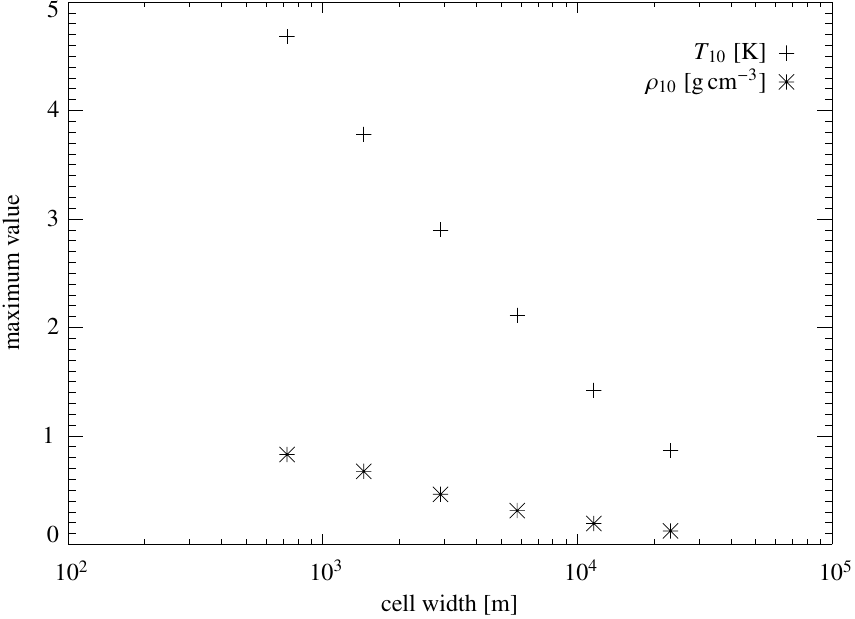}}
  \caption{One-dimensional resolution study of the maxima in the
    central shock collision of model s4.10A.}
  \label{fig:resstudy2}
\end{figure}

The numerical resolution of the parameter study presented above was
relatively low with a spatial resolution of $23.2~\mathrm{km}$ only.
Hence, for one of the most interesting models, z4.24A\_2d\_256, we
performed a resolution study.  Globally, we found a good agreement of
the temporal evolution of the state variables.  But the maximum
densities and temperatures at the central shocks collision showed a
significant dependence on the numerical resolution of the simulation
(see Fig.~\ref{fig:resstudy}).
With decreasing cell size the maximum values increase almost
exponentially.  This trend continues down to the smallest tested cell
width (7.4~km) and no indication for numerical convergence is found.
Given that the shock collision is an almost perfect spherically
symmetric problem, we reduced to one dimension in order to facilitate
even higher resolutions.  The state variables on the $r$-axis of
simulation s4.10A\_2dq\_256 at $t = 0.25~\mathrm{s}$ linearly
interpolated to higher resolutions were taken as initial condition.
The results of the corresponding one-dimensional runs with increasing
resolution are presented in Fig.~\ref{fig:resstudy2}.
It shows an analogous exponential increase of the temperature and
density maxima down to a cell width of 0.72~km.  Below that value,
however, the temperatures exceeded the limits of our equation of
state.  Here, more micro-physical processes would have to be taken
into account and could restrict further growth of density and
temperature.

But even without this restriction, from analytical considerations it
is clear that the density growth ratio must be limited, although it
can get very big.  In the idealized case of infinitely strong shock
reflection at the origin in spherical geometry (and constant initial
conditions of $\rho_0(r) = 1$, $u_0(r) = -1$, and $p_0(r) = 0$), its
value is $\left(\frac{\gamma + 1}{\gamma - 1}\right)^3$ after the
reflected shock \citep[cf.][]{Glaister1988}.\footnote{For an
  ultra-relativistic degenerate gas like in a WD's center $\gamma =
  \frac{4}{3}$ holds in good approximation and thus
  $\left(\frac{\gamma + 1}{\gamma - 1}\right)^3 = 7^3 = 343$.}
Another problem that can be solved semi-analytically is that of a
spherical shock wave that starts from infinity, converges to the
origin, and is reflected there
\citep[cf.][]{Guderley1942,Landau1991,Ponchaut2006}.  While the shock
is moving inwards it grows in strength because of its surface
decrease.  Close enough to the center for the strong shock
approximation to hold, it can be shown that $p \sim
r^{-\frac{2(1-n)}{n}}$, $\rho = \mathrm{const}$, and $|u| \sim
r^{-\frac{1-n}{n}}$ with $n$ being a constant between 0 and 1 that has
to be numerically determined and depends only on the ratio of specific
heats $\gamma$.  According to \citet{Landau1991}, at the time of
perfect focusing the density has increased by a factor of 20.1 for
$\gamma = \frac{7}{5}$ and the reflection is accompanied by another
increase by about a factor of 145.  Thus a maximum compression of
about 2,900 can be achieved and in the case of $\gamma = \frac{4}{3}$,
relevant in our case, those numbers are even bigger.

In our simulations the shock that propagates inwards into the WD due
to the He detonation is followed by an extended wave of matter that is
also flowing inwards.  Therefore the case might lie somewhere in
between these two idealized situations: The shock is amplified by the
geometrical focusing effect and thus compresses a significant part of
the core that is then burned at higher densities.  On the other hand
the matter succeeding the in-flowing shock leads to even higher
densities after the shock is reflected.  Taking all those properties
into account, it becomes clear why the ignition conditions for a core
detonation can be fulfilled so well in our simulations.  Provided that
the shock is symmetric enough also an off-center ignition at a point
with much smaller initial density like in model z4.24A\_2d\_256 is
possible in this scenario.

As the minimum shock surface that we can reach depends on the grid
resolution, the maximum temperature and density values that are
achieved through the geometric amplification in the shock implosion
naturally also do so.  In our simulations a successful second
detonation ignition thus depends on the resolution and it seems likely
that all the models in Table~\ref{tab:flmod} would explode provided
that they were well enough resolved.

Therefore, the results given in Table~\ref{tab:flmod} have to be
interpreted as a conservative estimate of the possibility of a
detonation ignition.  If at a certain grid resolution the ignition
conditions are fulfilled, a second core detonation must happen.
However, if the ignition conditions are not fulfilled a core
detonation can still not be excluded and might be achieved with
increasing resolution\footnote{Note that in this case the minimum
  detonator volume may be the limiting constraint.}.

\subsection{Three double-detonation supernova simulations}
\label{sec:dds}

As a first step to compare the simulations of sub-Chan\-dra\-se\-khar
mass models with actual SN~Ia observations, measurable quantities like
the nucleosynthetic abundances of an explosion have to be determined.
To this end, two complete double detonations of model z4.24A until $t
= 2.0~\mathrm{s}$ have been simulated.  One of them was set up in 3D
with $256^3$ cells for the full WD\@.  For comparison with previous
results, the second, 2D rotationally symmetric simulation was also
done with this low resolution ($128 \times 256$ cells).

\begin{figure}
  \resizebox{\hsize}{!}{\includegraphics{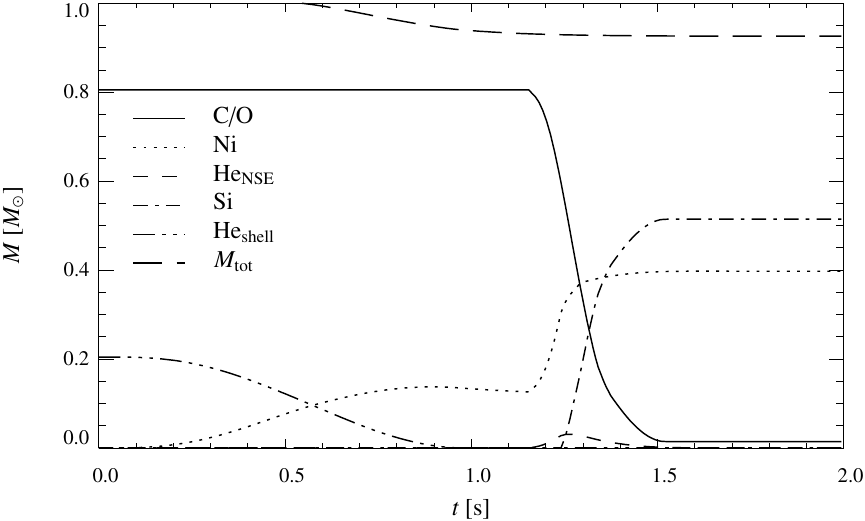}}
  \caption{Time evolution of the mass fractions of model
    z4.24A\_2d\_128.}
  \label{fig:mass}
\end{figure}

In order to test the influence of symmetry in the initial flame
configuration in the He shell on the explosion abundances, besides
these maximally asymmetric cases a 2D double detonation of the
spherically symmetric model s4.10A was also simulated.  The results of
all three simulations are given in Table~\ref{tab:dd}.
\begin{table}
  \caption{Results of three complete double detonation simulations.
    All nucleosynthetic abundances are given in solar masses.
    $M_\mathrm{loss}$ is the mass loss over the grid boundaries.}
  \label{tab:dd}
  \centering
  \begin{tabular}{lcccc}
    \hline
    \hline
    Model & \ncarb+\nox & \nni & \nsi & $M_\mathrm{loss}$ \\
    \hline
    z4.24A\_2d\_128 & 0.01 & 0.40 & 0.51 & 0.08 \\
    z4.24A\_3d\_128 & 0.01 & 0.45 & 0.50 & 0.05 \\
    s4.10A\_2dq\_128 & 0.02 & 0.41 & 0.50 & 0.08 \\
    \hline
  \end{tabular}
\end{table}
For model z4.24A\_2d\_128 also the time evolution of the mass
fractions of the species is plotted in Fig.~\ref{fig:mass}
and will be discussed in the following.

At first, the \nhe\ in the shell is burnt completely to \nni.  But the
mass of the latter is somewhat reduced due to some mass loss through
the boundaries of the computational domain.  This is clearly visible
also in the decrease of the total mass.  The reason for this is that
the co-expanding grid was set up to track a mass shell at
$0.85~\msol$.  Since the matter above this shell expands with a larger
velocity, some of it leaves the numerical grid.  The tracking of a
bigger mass shell would reduce this mass loss, but it would also
reduce the resolution of the colliding shocks in the core.

After about $1.2~\mathrm{s}$ the collision of the shock waves
resulting from the He detonation gives rise to conditions for a
detonation in the C~+~O core.  In the corresponding grid zones, this
detonation was artificially triggered by setting up a second level
set.  At the high densities near the center the C/O mixture is burnt
completely to \nni\ in NSE with a small amount of \nhe.  But the \nhe\ 
recombines to \nni\ after a short time due to the density decrease
caused by expansion.  Farther away from the center at smaller
densities the burning product is \nsi\ and some low-density matter at
the border of the core remains unburnt.  In this way $0.40~\msol$ of
\nni\ and $0.51~\msol$ of \nsi\ are produced after about
$1.5~\mathrm{s}$ and $0.01~\msol$ of C/O remain unburnt.

The results of the 3D case z4.24A\_3d\_128 are very similar to the 2D
one.  However, $0.05~\msol$ more \nni\ were produced.  The main reason
for this difference is the loss of different amounts of mass over the
boundaries of the computational domain, which amounted to $0.08~\msol$
in 2D and only $0.05~\msol$ in 3D\@.  The explanation for the smaller
mass loss is a simple geometric effect: In 3D cartesian coordinates
there is more volume outside the star that lies on the numerical grid
than in 2D cylinder coordinates.  For example, the largest distance
from the center is the length of the space diagonal ($\sqrt{3}\,a$,
when $a$ is the side length), whereas in 2D it is the length of the
diagonal in the $rz$-plane ($\sqrt{2}\,a$).

Due to the spherical symmetry in the model s4.10A\_2dq\_128 He burning
and shock convergence are faster.  Therefore all the reactions finish
already after about one second, but according to Table~\ref{tab:dd},
almost the same nucleosynthetic abundances are produced.  This
indicates that the density distribution in the core in case of shocks
converging asymmetric to the density distribution differs only
slightly from the perfect spherical symmetric case.  As the now
following core detonation propagates supersonic, the final abundances
are independent of the position of the ignition point.

In all three cases considered the \nni-mass and with it the explosion
energy are rather large.  In fact, they reach the range of (weaker)
\emph{normal} SNe~Ia.  However, these numbers have to be taken with
care since the density limit of $3 \cdot 10^7~\gqcm$ adopted here for
separating burning of C/O to \nni\ and \nsi, respectively, is quite
uncertain, but will have an impact on the final abundances.

\begin{figure}
  \resizebox{\hsize}{!}{\includegraphics{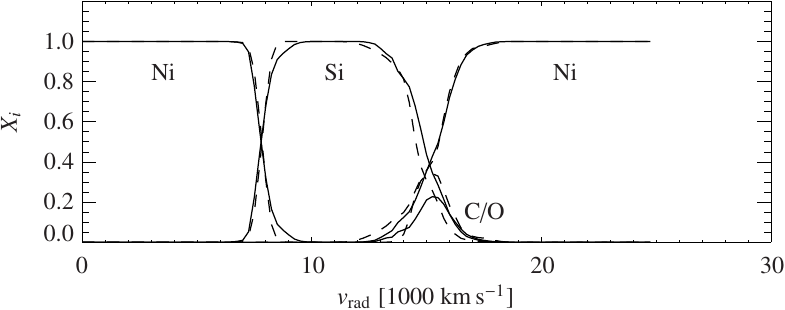}}
  \caption{Mass fractions of the species (averaged over all
    angles) of models z4.24A\_2d\_128 (solid line) and
    s4.10A\_2dq\_128 (dashed line) in radial velocity space at $t =
    10~\mathrm{s}$.}
  \label{fig:velprofzs}
\end{figure}

We could not calculate synthetic spectra or light curves of our
simulations yet.  First order spectra and light curves are given,
e.g., in \citet{Livne1995} for models very similar to our z4.24A case.
For comparison with their results the distributions of the species of
models z4.24A\_2d\_128 and s4.10A\_2dq\_128 in radial velocity space
(averaged over all angles) are presented in Fig.~\ref{fig:velprofzs}.
Both are very similar except for a minor difference in the unburnt
C/O, which is more abundant in the spherically symmetric case.  The
velocity ranges are comparable to model 6 of \citet{Livne1995}, which
had a similar setup ($M_\mathrm{C/O} = 0.8~\msol$, $M_\mathrm{He} =
0.17~\msol$), but resulted in a larger \nni\ mass ($0.648~\msol$).
Also, the maximum velocity of \nni\ and the minimum velocity of \nsi\ 
from core burning are shifted to significantly larger values.  This
could be caused by a difference in the transition density to NSE that
might also be responsible for the difference in \nni\ mass.
  
\begin{figure}
  \resizebox{\hsize}{!}{\includegraphics{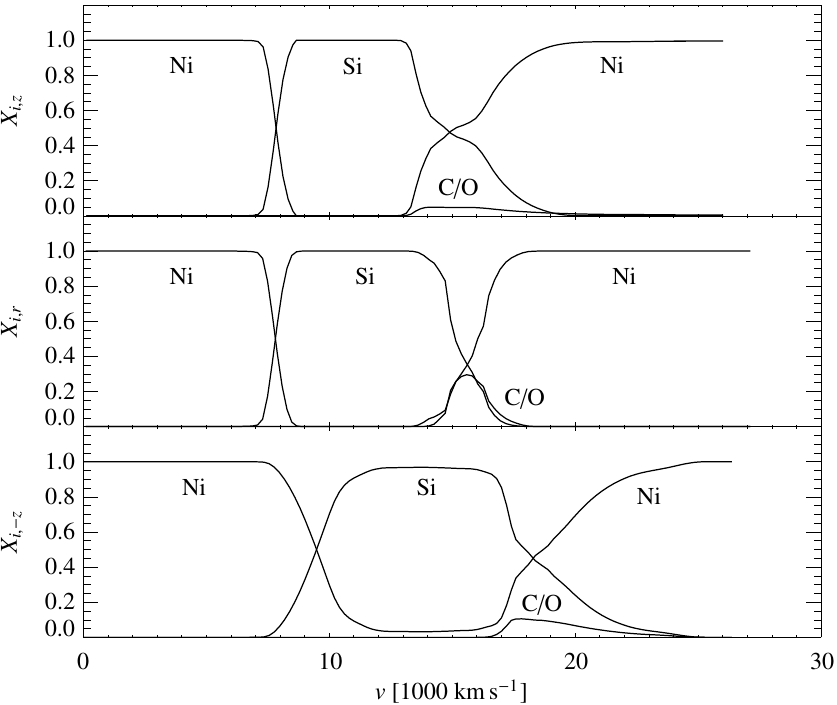}}
   \caption{Mass fractions of the species of model
     z4.24A\_2d\_128 in velocity space along the $z$-, $r$-, and
     negative $z$-direction at $t = 10~\mathrm{s}$.}
  \label{fig:velprofasym}
\end{figure}

In order to test possible asphericities of the ejecta in model
z4.24A\_2d\_128 the abundances are plotted separately for the $z$-,
$r$-, and $-z$-directions in Fig.~\ref{fig:velprofasym}.  Despite the
off-center explosion, the differences in the velocity distribution are
minor.  Only in $-z$-direction a shift in the maximum velocities of
the \nni\ from the core detonation and of the \nsi\ is visible.

\section{Summary and conclusions}
\label{sec:con}

We have studied the sub-Chandrasekhar model of SNe~Ia by means of a
series of two-dimensional and a few three-dimensional simulations with
different initial conditions.  The numerical scheme used is based on
the PPM algorithm, and the propagation of the detonation front is
modeled applying the level set technique.  This novel implementation
allowed for an accurate treatment of the hydrodynamic features of the
sub-Chandrasekhar model (such as shock waves and thin detonation
fronts) in multiple dimensions.

With our generic 1-\msol\ model we performed a parameter study
involving different He masses and ignition geometries of the initial
He detonation.  We find that a detonation in the He shell can clearly
trigger a second detonation in the core and at least for the mass
configurations studied, the double detonation seems to be a very
robust process, which works without any ``fine-tuning'' of our model
parameters.  In almost all of the simulations performed, the ignition
conditions for a core detonation were reached at or near the center of
the WD as a result of the convergence of the shock from the He shell
detonation and in the few other cases the failure is likely a result
of the finite numerical resolution.  The high maximum densities and
temperatures that are needed for detonation ignition and the fact that
the shock is not diluted too much when propagating inwards against the
pressure gradient are made possible through a geometrical shock
amplification effect that appears in spherical symmetry: As the
surface of the shock decreases its strength has to increase.  The high
compression that can be achieved theoretically in this way, especially
if full convergence is reached and the shock is reflected (cf.\ 
Sect.~\ref{sec:res}), could even allow successful double detonations
with considerably smaller He masses.  The shock amplification that is
reached, however, turns out to be resolution dependent on the
numerical grid as it is coupled to the smallest resolvable shock
surface.  Taking this into account it is very likely that all our
models would explode, if they were simulated with sufficiently high
resolution.

The question of whether an incineration could also happen directly at
the core--shell interface, was not addressed here.  This could of
course prevent the spherical shock convergence mechanism from playing
a role, at least in parts of the parameter space.  Therefore, edge-lit
detonations are postponed to a separate study.

The complete double detonation simulations that were performed for the
models z4.24A and s4.10A resulted in \nni-masses of about $0.40$ to
$0.45~\msol$.  Are the studied events thus good candidates for normal
SNe~Ia?  Most likely not.  Starting with a He-shell detonation, all of
our models show a significant amount of rapidly expanding \nni\ in the
outer layers.  In the observed spectra of normal and sub-luminous
events, however, this has never been observed (cf.\ 
\citet{Branch1982,Branch1984a,Woosley1986,Arnett1997} and also the
discussion in \citet{Livne1995}).  The only exception is the
super-luminous \object{SN~1991T}.  Thus our solar-mass models most
likely will not be able to explain normal or sub-luminous SNe~Ia and,
given the robustness of the explosion mechanism of the model, the only
conceivable explanation for the lack of observations of corresponding
SN~Ia events is that the progenitors are not realized in nature -- or
are very rare.

For a better agreement with observations a reduction of the He-shell
mass would be the most obvious choice.  However, then a significantly
larger core mass would most likely be required, because otherwise the
He would not detonate.  In this case the core density would be much
higher resulting in a very large Ni mass.  This would again not be a
candidate for normal SNe~Ia, but it might be a promising candidate for
objects like \object{SN~1991T}.

Further work could cover model improvements like a more realistic
treatment of nuclear reactions including the calculation of real
reaction rates depending on the actual thermodynamic state of the
burnt matter.  This would make an investigation of the onset and the
explosion dynamics of the core detonation possible.  Also, more
extended parameter studies, especially towards the lower end of
possible He masses, would be interesting.  For a comparison with
observations the calculation of synthetic light curves and spectra
would also be desirable.

\begin{acknowledgements}
  We want to thank Konstantinos Kifonidis who helped with the
  implementation of the Marquina solver and Ewald M\"uller for many
  helpful discussions.  Moreover we want to thank Eli Livne for very
  constructive comments on the first version of this paper.  The
  simulations presented here were carried out at the Computer Center
  of the Max Planck Society, Garching, Germany.  This work was
  supported by the Deutsche Forschungsgemeinschaft via the
  Transregional Collaborative Research Center TRR~33 ``The Dark
  Universe''.
\end{acknowledgements}

\bibliographystyle{aa} \bibliography{lit}

\end{document}